\documentclass[pra,fleqn,showpacs,floatfix,onecolumn]{revtex4}

\usepackage{hyperref}
\usepackage{rotating}
\usepackage{amsmath}

\begin{document}


\title{H$_2^+$ and HD$^+$: candidates for a molecular clock}


\author{J.-Ph. Karr}
\email{karr@lkb.upmc.fr}

\affiliation{Universit\'e d'Evry-Val d'Essonne, Boulevard Fran\c cois Mitterrand, 91025 Evry Cedex, France}
\affiliation{Laboratoire Kastler Brossel, UPMC-Paris 6, ENS, CNRS ; Case 74, 4 place Jussieu, 75005 Paris, France}

\begin{abstract}
We investigate the leading systematic effects in ro-vibrational spectroscopy of the molecular hydrogen ions H$_2^+$ and HD$^+$, in order to assess their potential for the realization of optical clocks that would be sensitive to possible variations of the proton-to-electron mass ratio. Both two-photon (2E1) and quadrupole (E2) transitions are considered. In view of the weakness of these transitions, most attention is devoted to the light shift induced by the probe laser, which we express as a function of the transition amplitude, differential dynamic polarizability and clock interrogation times. Transition amplitudes and dynamic polarizabilites including the effect of hyperfine structure are then calculated in a full three-body approach to get a precise evaluation of the light shift. Together with the quadrupole and Zeeman shifts that are obtained from previous works, these results provide a realistic estimate of the achievable accuracy. We show that the lightshift is the main limiting factor in the case of two-photon transitions, both in H$_2^+$ and HD$^+$, leading to expected accuracy levels close to $5 \times 10^{-16}$ in the best cases. Quadrupole transitions have even more promising properties and may allow reaching or going beyond $1 \times 10^{-16}$.
\end{abstract}




\maketitle

\section{Introduction}

Space-time variations of the proton-to-electron mass ratio $\mu$ is one area where high-precision molecular spectroscopy plays an important role in testing the fundamental laws of physics. While astrophysical studies on the spectra of molecular hydrogen, ammonia and methanol have set stringent limits on the variations of $\mu$ over cosmological time scales, laboratory tests provide complementary information on its variations in the current epoch (see~\cite{Jansen2014,Kozlov2013} for recent reviews). So far, the most severe model-independent limit from a laboratory measurement, obtained by comparing rovibrational transitions in SF$_6$ with a Cs clock, is at a level of several 10$^{-14}$/year~\cite{Shelkovnikov2008}. A large number of theoretical proposals relying on molecular lines with greatly enhanced sensitivities to $\mu$ variations, due to a cancellation between energy contributions from different origins (electronic, vibrational, rotational, fine or hyperfine structure...), have been formulated in recent years (see e.g.~\cite{Flambaum2007,DeMille2008,Kajita2011}).

Ro-vibrational transitions in the hydrogen molecular ions H$_2^+$ and HD$^+$ were among the very first candidates to be studied in the perspective of high-precision laboratory tests~\cite{Schiller2005}. Despite the fact that they do not benefit from any sensitivity enhancement effect, several factors still make hydrogen molecular ions attractive. First are the small natural widths of ro-vibrational lines, which lye in the 10~Hz~\cite{Peek1979,Pilon2013} and 10$^{-7}$~Hz~\cite{Pilon2012} range for HD$^+$ and H$_2^+$ respectively, and the possibility to probe single trapped molecular ions with ultra-high resolution using quantum-logic spectroscopy (QLS)~\cite{Schmidt2005,Mur-Petit2012}. The demonstration of efficient schemes to prepare the molecules in a selected ro-vibrational state, such as resonance-enhanced multi-photon ionization~\cite{Pratt1995,OHalloran1987} or rotational cooling~\cite{Staanum2010,Schneider2010}, is another important asset; in general, state preparation is a major experimental challenge to be considered since transitions of enhanced sensitivity are often rather exotic, i.e. involving highly excited states in complex molecules. Finally, the simplicity of hydrogen molecular ions allows for precise theoretical calculations of the main systematic shifts that affect the accuracy of the measurement, and thus reliable performance estimates of the envisaged quantum-logic clock. The purpose of the present work is to obtain such estimates for a few ro-vibrational transitions of interest in H$_2^+$ and HD$^+$.

This study builds upon a previous investigation of systematic effects for a large number of transitions in HD$^+$, by D. Bakalov and S. Schiller~\cite{Bakalov2013}, which included the combined contributions of Zeeman, quadrupole, Stark, and black-body radiation shifts, and concluded that frequency measurements with a relative accuracy of about $5 \times 10^{-16}$ should be achievable on a few selected one-photon ro-vibrational transitions. Two-photon transitions, which enable suppression of the first-order Doppler effect even without strong spatial confinement of the ions, were also considered (see also~\cite{Tran2013}). However, one of the potentially most important systematic effects that is the light shift induced by the probe laser, was not calculated. The first Section is devoted to a precise estimate of this effect, and consideration is extended to the H$_2^+$ ion. One attractive feature of H$_2^+$ in this context is the extremely small natural width of the transitions, to which one might add the simpler spin structure. In the second Section, we study one-photon quadrupole transitions in H$_2^+$, which benefit from smaller light shifts. It is shown that these transitions could allow for even more accurate frequency measurements, and are thus highly promising candidates for a molecular ion clock.

\section{Two-photon transitions}

\subsection{Expression of the light shift} \label{ls-deriv}

In order to give a correct estimate the light shift, one should first discuss the typical intensity of the probe laser to be used for optimal operation of a molecular ion clock. Here, the envisaged experimental realization is QLS of a Be$^+$/HD$^+$ (or Be$^+$/H$_2^+$) ion pair. The use of two-photon transitions implies reaching a regime of two-photon Rabi oscillations; the generalized two-photon Rabi frequency for a transition between states $i$ and $f$ is given by
\begin{equation}
\Omega = \frac{\left| ^SQ_{\epsilon \epsilon}^{if}(E) \right| I}{\epsilon_0 \hbar c}
\end{equation}
where $I$, $\epsilon$ are the intensity and polarization of the probe laser, and $^SQ_{\epsilon_1 \epsilon_2}^{if} (E) = \langle i | ^SQ_{\epsilon_1 \epsilon_2} (E) | f \rangle$ the two-photon transition matrix element. $E = (E_i + E_f)/2$ is the intermediate energy, and the two-photon operator $^SQ_{\epsilon_1 \epsilon_2} (E)$ is defined by
\begin{equation}
^SQ_{\epsilon_1 \epsilon_2} = \frac{1}{2} \left( Q_{\epsilon_1 \epsilon_2} + Q_{\epsilon_2 \epsilon_1} \right), \hspace{3mm} Q_{\epsilon_1 \epsilon_2} = \mathbf{d} \cdot \boldmath{\epsilon_1} \frac{1}{E-H} \mathbf{d} \cdot \boldmath{\epsilon_2}. \label{two-photon-op}
\end{equation}
QLS clocks are operated with a single interrogation $\pi$-pulse~\cite{Rosenband2007} whose duration $\tau$ is adjusted according to the desired resolution. $\tau$ is typically of the order of 100~ms~\cite{Rosenband2008,Chou2010}, and we will assume the same value in the case of H$_2^+$. However, in HD$^+$ the shorter lifetime $\tau_f$ of excited ro-vibrational state ($\tau_f \sim 15-55$~ms for $v=1-4$) constrains $\tau$ to lower values; in the following we will assume $\tau = \tau_f/10$. The probe laser intensity is determined from the relationship $\Omega \tau = \pi$ yielding
\begin{equation}
I = \frac{ \pi \epsilon_0 \hbar c}{\tau \left| ^SQ_{\epsilon \epsilon}^{if} \right|}
\end{equation}
The corresponding shift of the transition energy $(E_f - E_i)/2$ is
\begin{equation}
\Delta E^{if} = - \frac{1}{4} \frac{\Delta \alpha_{\epsilon}^{if} \; I}{c} = \frac{\pi \epsilon_0 \hbar} {4 \tau} \frac{\Delta \alpha_{\epsilon}^{if}}{\left| ^SQ_{\epsilon \epsilon} ^{if}(E) \right|},
\end{equation}
where $\Delta \alpha_{\epsilon}^{if} = \alpha^f_{\epsilon}(\omega_{if}) - \alpha^i_{\epsilon} (\omega_{if})$, $\omega_{if} = (E_f - E_i)/2\hbar$ being the two-photon probe laser frequency and $\alpha^n_{\epsilon} (\omega)$ the dynamical polarizability of a state $n$:
\begin{equation}
\alpha^n_{\epsilon} (\omega) = -4\pi a_0^3 \left( ^SQ_{\epsilon \epsilon} ^{nn} (E_n + \hbar\omega) + ^SQ_{\epsilon \epsilon} ^{nn}(E_n - \hbar\omega) \right). \label{polarizability}
\end{equation}
Using the atomic units for polarizabilites and two-photon transition matrix elements (i.e. $\alpha^n_{\epsilon}$ in units of $4\pi a_0^3$ and $^SQ_{\epsilon_1 \epsilon_2}^{if}$ in units of $(ea_0)^2/E_h = \hbar^2 e^2 / m_e$) one obtains the following simple expression for the frequency shift:
\begin{equation}
\Delta f_{LS} = -\frac{1}{8\tau} \; \frac{\Delta \alpha_{\epsilon}^{if}}{\left| ^SQ_{\epsilon \epsilon} ^{if}(E) \right|}. \label{lightshift}
\end{equation}
In order to get a reliable estimate of the light shift, we have calculated the dimensionless transition amplitudes $\left| ^SQ_{\epsilon \epsilon}^{if}(E) \right|$ and associated differential polarizabilities $\Delta \alpha_{\epsilon}^{if}$ for a large range of two-photon transitions in HD$^+$ and H$_2^+$. Selected results are presented in Tables~\ref{lshdp} and \ref{lsh2p}. The methods for calculating two-photon line strengths with account of the hyperfine structure have been described in detail in~\cite{Karr2008a} in the context of H$_2^+$ (see also~\cite{Bakalov2005,Bakalov2012} for similar calculations in HD$^+$); they can also be applied for evaluating dynamical polarizabilities of hyperfine sublevels since both quantities are expressed in terms of matrix elements of the same operator $^SQ_{\epsilon_1 \epsilon_2} (E)$. A summary of these methods can be found in the Appendix.

\subsection{HD$^+$}

In the case of HD$^+$ (Table~\ref{lshdp}), two-photon transitions from ro-vibrational states $(v=0,L)$ were considered, since $v=0$ states, having by far the longest lifetimes, are the only ones in which the ions can be conveniently prepared. We mainly investigated the transitions towards states with a vibration quantum number $v'=2$ or $1$ which are the strongest, as shown in~\cite{Karr2005}. Indeed, in a $(v=0,L) \to (v'=2,L')$ transition, a $(v''=1,L'')$ state lies close to the one-photon resonance, leading to a strong enhancement of the line strength.  Transitions towards higher vibrational states are much weaker and more difficult to observe experimentally; in view of Eq.~(\ref{lightshift}) they will also be affected by large lightshifts unless the dynamical polarizabilities of the ground and excited states cancel each other almost exactly. We also investigated $v=0 \to v=4$ transitions but found no such cancellations.

The angular momentum coupling scheme used to denote the hyperfine sublevels $(v,L,F,S,J,J_z)$ of HD$^+$ is the following~\cite{Bakalov2006}: $\mathbf{F} = \mathbf{S_d} + \mathbf{S_e}$, $\mathbf{S} = \mathbf{F} + \mathbf{S_p}$, $\mathbf{J} = \mathbf{L} + \mathbf{S}$ where $\mathbf{S_i}$, $i = e,p,d$ denote the spin operator of the electron, proton and deuteron; only $J$ is an exact quantum number whereas $(F,S)$ are approximate quantum numbers. As discussed in~\cite{Bakalov2011,Bakalov2013}, two types of transitions are particularly interesting for clock applications:
\begin{itemize}
\item $J_z = 0 \to J'_z = 0$ transitions, which only have a quadratic Zeeman shift;
\item pairs of transitions between ''stretched states'' where all the angular momenta have their maximum values: $(v,L,F=1,S=2,J=L+2,J_z = \pm(L+2)) \to (v',L',F'=1,S'=2,J'=L'+2,J'_z = \pm(L'+2))$. Since these transitions have strictly linear opposite Zeeman shifts, the effect of the magnetic field can be nulled by measuring both transition frequencies and computing the average value.
\end{itemize}
A few $J_z = 0 \to J'_z = 0$ transitions benefit from a relatively low Zeeman shift (below 100 Hz in a magnetic field of 1 G): $(v=0,L=0) \to (v'=2,L=0)$ with $(F,S,J) = (0,1,1)$, $(v=0,L=3) \to (v=2,L=3)$ with $(F,S,J) = (1,2,4)$, and $(v=0,L=4) \to (v=2,L=4)$ with $(F,S,J) = (1,2,3)$ (see Table 2 of~\cite{Bakalov2011}). However, none of them were found to have interesting properties in terms of systematic effects, and they will not be considered further. In the following we only discuss transitions between stretched states.

The first seven lines of Table~\ref{lshdp} give the lightshifts of all $v=0 \to v=2$ stretched-state transitions with an initial rotational state $L \leq 2$. $v=0 \to v=2$ transitions are obvious candidates since they are the most intense, and the first three of them were considered in~\cite{Bakalov2013} (Table IX). Their lightshifts range from a few to a few hundred Hertz, the most promising among them being the $(v=0,L=0) \to (v'=2,L'=0)$ transition which has a 4.0~Hz light shift and is not affected by Zeeman and quadrupole shifts. The relative frequency shift amounts to $\Delta f_{LS}/f_0 = 7 \times 10^{-14}$ and is the main limitation to the accuracy. Assuming that it could be calibrated to the percent level, a residual systematic uncertainty of $7 \times 10^{-16}$ could be achieved; lightshift calibration beyond that level would be a difficult experimental challenge. The $(v=0,L=1) \to (v'=2,L'=1)$ transition, which seemed promising due to its large intensity as pointed out in~\cite{Karr2005}, turns out to have a large lightshift, partly due to the weakness of the Zeeman sub-lines which connect the stretched states. We extended our search to $L \leq 5$ initial states and found one more interesting transition (line before last in Table~\ref{lshdp}): $(v=0,L=4) \to (v'=2,L'=2)$, which has both a moderate light shift ($\Delta f_{LS}/f_0 = 1.9 \times 10^{-13}$) and low quadrupole shift ($\Delta f_Q/f_0 = 3.5 \times 10^{-15}$).

The $v=0 \to v=1$ transitions, although less intense, may also be of interest in case of a fortuitous cancellation of the lightshift. A scan of all transitions from $(v=0,L \leq 5)$ states revealed one such instance (last line of Table~\ref{lshdp}): $(v=0,L=2) \to (v'=1,L'=0)$, with a lightshift of 0.34~Hz ($\Delta f_{LS}/f_0 = 1.3 \times 10^{-14}$). For this transition, the accuracy would be limited by the quadrupole shift ($\Delta f_Q/f_0 = 2 \times 10^{-13}$), which according to the discussion in Ref.~\cite{Bakalov2013} could be reduced to a residual uncertainty of about $1 \times 10^{-15}$ by applying a nulling procedure.

\begin{sidewaystable}
\begin{tiny}
\begin{tabular}{|c|c|c|c|c|c|c|c|c|c|c|c|c|c|c|c|c|c|}
 \hline
$(v,L)$ & $(v',L')$ &    $f_0$    & $f_0$ & $\tau_f$ & $Q_{v,L,v',L'}^{(0)}$ & $Q_{v,L,v',L'}^{(2)}$ & $|Q_{if}|$ & $\alpha_{v,L}^{(0)}$ & $\alpha_{v,L}^{(2)}$ & $\alpha_i$ & $\alpha_{v',L'}^{(0)}$ & $\alpha_{v',L'}^{(2)}$ & $\alpha_f$ & $\Delta\alpha$ & $\Delta f_{LS}$ & $\Delta f_Z$ & $\Delta f_Q$
     \\
&                   & [cm$^{-1}$] & [THz] &  [ms]    &   [a.u.]              &     [a.u.]            &   [a.u.]   &       [a.u.]         &        [a.u.]        & [a.u.]     &    [a.u.]              &     [a.u.]             &   [a.u.]   &      [a.u.]    &     [Hz]        &    [Hz]      &   [Hz]
     \\
 \hline
(0,0) & (2,0) & 1864.9281 & 55.91 & 29.3 & 1.4822 & 0        & 1.4822  & 3.8681 & 0        & 3.8681 & 4.0086  & 0        & 4.0086  & 0.1405  & 4.0   & 0         & 0
\\
(0,0) & (2,2) & 1924.6823 & 57.70 & 30.4 & 0      & 7.6132   & 4.1699  & 5.7365 & 0        & 5.7365 & 5.1294  & -5.2596  & 6.5351  & 0.7986  & 7.9   & $\mp$549  & 7.1
\\
(0,1) & (2,1) & 1862.9561 & 55.85 & 29.5 & 7.9307 & -23.1605 & 0.6067  & 8.7670 & -19.1522 & 2.7105 & 11.4708 & -29.6063 & 2.1085  & -0.6021 & -42.0 & $\pm$5    & 1.0
\\
(0,1) & (2,3) & 1962.1424 & 58.82 & 30.6 & 0      & 4.9846   & 3.9966  & 4.1738 & -1.1584  & 4.3570 & 5.1502  & -4.4530  & 6.5874  & 2.2304  & 22.8  &  $\mp$544 & 4.4
\\
(0,2) & (2,0) & 1799.2675 & 53.94 & 29.3 & 0      & 5.5963   & 6.8540  & 4.9353 & 2.7023   & 4.2131 & 5.6796  & 0        & 5.6796  & 1.4665  & 9.1   &  $\pm$558 & -5.6
\\
(0,2) & (2,2) & 1859.0217 & 55.73 & 30.4 & -1.0086 & 2.4984  & 0.3269  & 1.9452 & 0.6514   & 2.2934 & -1.8277 & 3.7284   & 0.1652  & -2.1282 & -268  &  $\pm$9   & 1.4
\\
(0,2) & (2,4) & 1997.0422 & 59.87 & 31.5 & 0      & 3.8963   & 3.5568  & 3.9275 & -1.7206  & 4.3820 & 5.1278  & -4.0079  & 6.5576  & 2.1756  & 24.3  & $\mp$538  & 3.5
\\
(0,4) & (2,2) & 1707.3273 & 51.18 & 30.4 & 0      & 9.2358   & 11.3115 & 5.8167 & -5.4879  & 7.7745 & 9.8934  & -2.4548  & 10.5495 & 2.7749  & 10.0  & $\pm$567  & -0.18
\\
(0,2) & (1,0) & 890.8371  & 26.71 & 54.6 & 0      & 0.3705   & 0.4538  & 2.2803 & -2.0531  & 2.8290 & 2.8223  & 0        & 2.8223  & -0.0067 & -0.34 & $\pm$558  & -5.6
\\
\hline
\end{tabular}
\caption{\label{lshdp} Light shifts of selected two-photon ro-vibrational transitions between stretched states $i=(v,L,F=1,S=2,J=L+2,J_z=\pm(L+2)) \to f=(v',L',F'=1,S'=2,J'=L'+2,J'_z=\pm(L'+2))$ in HD$^+$. $f_0 = (E_f-E_i)/2$ is the spin-independent transition frequency (including leading-order relativistic and QED corrections) taken from~\cite{Moss1993hdp}. $\tau_f$ is the excited state lifetime from~\cite{Pilon2013}. $Q_{v,L,v',L'}^{(k)}$ are the scalar ($k=0$) and tensor ($k=2$) two-photon matrix elements (see Eq.~(\ref{two-photon-def})), and $Q_{if}$ is the resulting amplitude for the hyperfine component that connects the stretched states $i,f$, obtained using Eq.~(\ref{polar-lin-lin}) or (\ref{polar-plus-plus}) according to the polarization case, and Eq.~(\ref{hfs-hdp}). Similarly, $\alpha_{v,L}^{(k)}$ and $\alpha_{v',L'}^{(k)}$ ($k=0,2$) are the scalar and tensor dynamic polarizabilities (Eq.~(\ref{polarizability})) of the initial and final ro-vibrational states, $\alpha_i$, $\alpha_f$ are the hyperfine state polarizabilities (obtained through Eq.~(\ref{polar-lin-lin}) or (\ref{polar-plus-minus}) with Eq.~(\ref{hfs-hdp})), and $\Delta\alpha = \alpha_f - \alpha_i$. A linearly (resp. circularly) polarized laser field was assumed for $L \to L$ (resp. $L \to L \pm 2$) transitions for which $|\Delta J_z| = 0$ (resp. 2). Finally, $\Delta f_{LS}$ is the estimated light shift, obtained from Eq.~(\ref{lightshift}). Estimates of the two other most important systematic effects are also given: the Zeeman shift $\Delta f_{Z}$ in a magnetic field of 1~G from~\cite{Bakalov2011}, and the quadrupole shift $\Delta f_{Q}$ in an electric field gradient of $0.67 \times 10^8$~V/m$^{2}$ (see~\cite{Bakalov2013} and corrigendum to be published).}
\end{tiny}
\end{sidewaystable}

\subsection{H$_2^+$}

In H$_2^+$ (Table~\ref{lsh2p}), two-photon transition amplitudes and polarizabilities follow a very different behaviour from the HD$^+$ case. Neighbouring rovibrational states not being dipole-allowed, there are no resonant enhancement effects. Only excited electronic states contribute to the two-photon amplitudes, leading to a smooth variation as a function of vibration and rotation quantum numbers. It has been shown that only $\Delta v =1$ two-photon transitions are sufficiently intense to be conveniently detected~\cite{Hilico2001}. In contradistinction with HD$^+$ the initial vibrational state could take any value (excited rovibrational states are metastable); however the selectivity in H$_2^+$ production by resonance-enhanced multiphoton ionization gets worse when the targeted vibrational quantum number increases~\cite{OHalloran1987}. $(v=0,L) \to (v'=1,L')$ and $(v=1,L) \to (v'=2,L')$ transitions with $L \leq 3$ are listed in Table~\ref{lsh2p}.

The spin structure of H$_2^+$ is slightly simpler than in HD$^+$~\cite{Korobov2006}. The adopted angular momentum coupling scheme is $\mathbf{F} = \mathbf{S_e} + \mathbf{I}$, $\mathbf{J} = \mathbf{L} + \mathbf{F}$ where $\mathbf{I}$ is the total nuclear spin, whose value is 0 (resp. 1) for even (resp. odd) values of $L$. No $J_z = 0 \to J'_z = 0$ transitions exist since the total angular momentum is half integer. There are however stretched states which have the same properties as in HD$^+$: $(v,L,F=1/2,J=L+1/2,J_z=\pm(L+1/2))$ for even $L$ and $(v,L,F=3/2,J=L+3/2,J_z=\pm(L+3/2))$ for odd $L$. In the following only transitions between such states are considered.

The light shifts lie in the few-Hz range, which is comparable to the best transitions in HD$^+$. In the absence of resonant enhancement effects, the $\Delta\alpha/|Q_{if}|$ ratio is less favorable, but this is compensated by the possibility of using longer probe times and thus lower intensities. $L \to L-2$ transitions (especially $L=2 \to L'=0$) have the lowest lightshifts, followed by $L \to L$ transitions, while $L \to L+2$ transitions are less favourable. The quadrupole shift has the same order of magnitude, and is lowest for $L \to L$ transitions. When $v$ is increased from 0 to 1, the lightshift slightly decreases but the quadrupole shift slightly increases. The best trade-off depends on the degree of accuracy to which the different systematic effects can be calibrated and corrected for. We will make the same assumptions as in Ref.~\cite{Bakalov2013} where this issue has been discussed in detail.

The Zeeman shift is nulled by computing the average of the stretched-state transitions $J_z \to J'_z$ and $-J_z \to -J'_z$. In~\cite{Bakalov2013} it was assumed that each transition frequency can be measured with a resolution equal to 1\% of the natural linewidth. In H$_2^+$ the natural width is extremely small, so that the transition width would probably be limited by the probe laser. For illustration let us assume a value of 1~Hz. Then the uncertainty on the mean frequency of the doublet would be $0.01 \times \sqrt{2} = 0.014$~Hz. By repeating these measurements for a set of magnetic field values the error could be reduced by a further factor of 5, i.e. slightly below 10$^{-16}$ relative to the transition frequency.

The quadrupole shift is nulled by measuring the transition frequencies for three orthogonal directions of the magnetic field. Again, the uncertainty of each measurement can be assumed to be 1\% of the linewidth, yielding an overall uncertainty of $0.01/\sqrt{3} = 0.0058$~Hz. In addition, the inaccuracy in establishing three perfectly orthogonal magnetic field directions leads to a residual uncertainty of 0.5\% of the value calculated in Table~\ref{lsh2p} for an electric field gradient of $0.67 \times 10^8$~V/m$^{2}$ (assuming this can be performed with the same accuracy as in the Hg$^+$ ion clock~\cite{Oskay2006}). Supposing, finally, that the residual uncertainty on the lightshift is equal to 1\% of the shift, and adding quadratically the different errors, the best trade-off would be obtained for $L \to L$ transitions, with overall accuracies between 5 and $7 \times 10^{-16}$ depending on the values of $v$ and $L$.

\begin{sidewaystable}
\begin{tiny}
\begin{tabular}{|c|c|c|c|c|c|c|c|c|c|c|c|c|c|c|c|c|c|}
 \hline
$(v,L)$ & $(v',L')$ &    $f_0$    & $f_0$ & $Q_{v,L,v',L'}^{(0)}$ & $Q_{v,L,v',L'}^{(2)}$ & $|Q_{if}|$ & $\alpha_{v,L}^{(0)}$ & $\alpha_{v,L}^{(2)}$ & $\alpha_i$ & $\alpha_{v',L'}^{(0)}$ & $\alpha_{v',L'}^{(2)}$ & $\alpha_f$ & $\Delta\alpha$ & $\Delta f_{LS}$ & $\Delta f_Z$ & $\Delta f_Q$
     \\
&                   & [cm$^{-1}$] & [THz] &   [a.u.]              &     [a.u.]            &   [a.u.]   &       [a.u.]         &        [a.u.]        & [a.u.]     &    [a.u.]              &     [a.u.]             &   [a.u.]   &      [a.u.]    &     [Hz]        &    [Hz]      &   [Hz]
     \\
 \hline
(0,0) & (1,0) & 1095.5633 & 32.84 & -0.4189 & 0      & 0.4189 & 3.1691 & 0       & 3.1691 & 3.8982  & 0        & 3.8982  & 0.7291 & 2.2  & 0         & 0
\\
(0,0) & (1,2) & 1178.0933 & 35.32 & 0       & 0.5536 & 0.3032 & 3.1692 & 0       & 3.1692 & 3.9360  & -2.0593  & 4.4864  & 1.3172 & 5.4  & $\mp$720  & 6.1
\\
(0,1) & (1,1) & 1094.0306 & 32.80 & -0.4204 & 0.3655 & 0.3048 & 3.1786 & -1.6939 & 2.6429 & 3.9106  & -2.4125  & 3.1477  & 0.5048 & 2.1  & $\pm$3.5  & 0.53
\\
(0,1) & (1,3) & 1230.8450 & 36.90 & 0       & 0.3917 & 0.3141 & 3.1788 & -1.6940 & 3.4466 & 3.9741  & -2.0189  & 4.6257  & 1.1791 & 4.7  & $\mp$690  & 3.4
\\
(0,2) & (1,0) & 1008.4449 & 30.23 & 0       & 0.2705 & 0.3313 & 3.1978 & -1.4455 & 3.5841 & 3.8981  & 0        & 3.8981  & 0.3140 & 1.2  &  $\pm$727 & -5.3
\\
(0,2) & (1,2) & 1090.9749 & 32.71 & -0.4234 & 0.3119 & 0.2567 & 3.1978 & -1.4455 & 2.4251 & 3.9359  & -2.0592  & 2.8352  & 0.4101 & 2.0  &  $\pm$7.2 & 0.76
\\
(0,2) & (1,4) & 1280.9902 & 38.40 & 0       & 0.3787 & 0.3457 & 3.1980 & -1.4456 & 3.5844 & 4.0252  & -2.0315  & 4.7499  & 1.1656 & 4.2  & $\mp$685  & 2.5
\\
(0,3) & (1,1) &  949.6011 & 28.47 & 0       & 0.3071 & 0.3761 & 3.2267 & -1.4169 & 3.6840 & 3.9104  & -2.4122  & 4.2918  & 0.6078 & 2.0  & $\pm$704  & -2.0
\\
(0,3) & (1,3) & 1086.4155 & 32.57 & -0.4280 & 0.3056 & 0.2307 & 3.2268 & -1.4169 & 2.3122 & 3.9739  & -2.0187  & 2.6708  & 0.3586 & 1.9  & $\pm$10.7 & 0.88
\\
\hline
(1,0) & (2,0) & 1031.9570 & 30.94 & -0.7004 & 0      & 0.7004 & 3.8981 & 0       & 3.8981 & 4.8223  & 0        & 4.8223  & 0.9242 & 1.6  & 0         & 0
\\
(1,0) & (2,2) & 1110.0556 & 33.28 & 0       & 0.9829 & 0.5384 & 3.8982 & 0       & 3.8982 & 4.8724  & -2.8800  & 5.6421  & 1.7439 & 4.0  & $\mp$712  & 6.9
\\
(1,1) & (2,1) & 1030.4768 & 30.89 & -0.7028 & 0.6495 & 0.4974 & 3.9105 & -2.4124 & 3.1476 & 4.8381  & -3.3710  & 3.7721  & 0.6245 & 1.6  & $\pm$3.8  & 0.57
\\
(1,1) & (2,3) & 1159.9357 & 34.77 & 0       & 0.7441 & 0.5966 & 3.9107 & -2.4126 & 4.2922 & 4.9232  & -2.8244  & 5.8348  & 1.5426 & 3.2  & $\mp$682  & 3.9
\\
(1,2) & (2,0) &  949.4269 & 28.46 & 0       & 0.4821 & 0.5904 & 3.9357 & -2.0590 & 4.4860 & 4.8220  & 0        & 4.8220  & 0.3360 & 0.71 & $\pm$720  & -6.1
\\
(1,2) & (2,2) & 1027.5256 & 30.80 & -0.7084 & 0.5547 & 0.4119 & 3.9358 & -2.0591 & 2.8352 & 4.8722  & -2.8797  & 3.3329  & 0.4978 & 1.5  &  $\pm$7.7 & 0.81
\\
(1,2) & (2,4) & 1207.3060 & 36.19 & 0       & 0.6761 & 0.6172 & 3.9361 & -2.0593 & 4.4865 & 4.9911  & -2.8432  & 6.0054  & 1.5189 & 3.1  & $\mp$677  & 2.8
\\
(1,3) & (2,1) &  893.6624 & 26.79 & 0       & 0.5679 & 0.6955 & 3.9737 & -2.0184 & 4.6251 & 4.8377  & -3.3706  & 5.3706  & 0.7455 & 1.3  & $\pm$697  & -2.3
\\
(1,3) & (2,3) & 1023.1213 & 30.67 & -0.7166 & 0.5439 & 0.3655 & 3.9738 & -2.0185 & 2.6709 & 4.9229  & -2.8242  & 3.0999  & 0.4290 & 1.5  & $\pm$11.7 & 0.95
\\
\hline
\end{tabular}
\caption{\label{lsh2p} Same as Table~\ref{lshdp}, for two-photon ro-vibrational transitions between stretched states $i=(v,L,F=3/2,J=L+2,J_z=\pm(L+2)) \to f=(v',L',F'=1,S'=2,J'=L'+2,J'_z=\pm(L'+2))$ in H$_2^+$. The spin-independent transition frequencies are taken from~\cite{Moss1993h2p}, and the Zeeman shift $\Delta f_{Z}$ is obtained from~\cite{Karr2008b}. The relationship between hyperfine matrix elements of the two-photon operator and its reduced matrix elements $Q_{v,L,v',L'}^{(i)}$ is given in Eq.~(\ref{hfs-h2p}).}
\end{tiny}
\end{sidewaystable}

\section{Quadrupole transitions in H$_2^+$}

In the H$_2^+$ ion, quadrupole transitions are a promising alternative as potential clock transitions. Their oscillator strengths have recently been calculated~\cite{Pilon2012}; they are more intense than two-photon transitions, so that the lighshift could be significantly reduced. In addition, the possibility to excite transitions towards higher vibrational states is attractive because increasing the transition frequency generally leads to improving clock performances. However, transitions become weaker when $\Delta v$ is increased, and again it is useful to estimate the lightshift in order to determine the $\Delta v$ range where it does not represent a limitation to the accuracy.

We will thus derive an expression of the light shift along the same lines as in Sec.~\ref{ls-deriv}. In the case of quadrupole transitions, the Rabi frequency writes~\cite{James1998}
\begin{equation}
\Omega = \frac{e a_0^2 E_0}{2\hbar} \; \frac{2\pi}{\lambda} \; |\Theta^{if}_{\epsilon,n}| \label{quad-rabi}
\end{equation}
with
\begin{equation}
\Theta^{if}_{\epsilon,n} = \frac{1}{3} \sum_{ij} \langle i | \Theta_{ij} | f \rangle \epsilon_i n_j \label{quad-with-polar}
\end{equation}
$E_0$, $\mathbf{\epsilon}$ are respectively the electric field amplitude and polarization of the incident wave, and $\mathbf{n}$ the unit vector along the propagation direction ($\mathbf{k} = k\mathbf{n}$). $\Theta$ is the quadrupole tensor operator in the center-of-mass frame in atomic units (i.e. units of $e a_0^2$). We use the definition of~\cite{Bakalov2013}:
\begin{equation}
\Theta_{ij} = (3/2) \sum_k Z_k \left(R_{ki} R_{kj} - \delta_{ij} \mathbf{R_k}^2/3 \right). \label{quad-def}
\end{equation}
where $\mathbf{R_k}$ is the position of the particle $k$ with respect to the center of mass, and $Z_k$ its electric charge. The condition $\Omega \tau = \pi$ (where $\tau$ is the probe pulse duration) yields
\begin{equation}
E_0 = \frac{\hbar \lambda}{e a_0^2 \tau |\Theta^{if}_{\epsilon,n}|},
\end{equation}
from which the corresponding lightshift is easily deduced:
\begin{equation}
\Delta f_{LS} = - \frac{m_e \lambda^2}{4 h} \; \frac{\Delta \alpha_{\epsilon}^{if}}{\tau^2 |\Theta^{if}_{\epsilon,n}|^2}.
\end{equation}
The choices of propagation direction and polarization of the probe laser, which influence both the quadrupole transition amplitude and lightshift, are given in the Appendix.

We calculated the dimensionless transition amplitudes $|\Theta^{if}_{\epsilon,n}|$ and associated differential polarizabilities $\Delta \alpha_{\epsilon}^{if}$ for $(v=0,L) \to (v',L')$ transitions with $v' = 1,2,3$. Our results are summarized in Table~\ref{lsquad}. $L \to L$ (with $L \leq 3$), $L=3 \to L'=1$ and $L=4 \to L'=2$ transitions were considered. $L \to L+2$ and $L =2 \to L' = 0$ transitions were omitted because they have larger quadrupole shifts as can be seen in Table~\ref{lsh2p}.

This study shows that the lightshift is negligible for $v=0 \to v'=1$ transitions; for $v=0 \to v'=2$ transitions, they are still significantly lower (by at least an order of magnitude) than in two-photon transitions. However, for $v=0 \to v'=3$ transitions they are already much higher. Under the same assumptions as above regarding calibration of the main systematic effect, we estimate that a relative accuracy of about $1-1.5 \times 10^{-16}$ could be reached using the transitions $(v=0,L) \to (v',L'=L)$ with $v'=1$ or 2. Especially promising is the transition $(v=0,L=4) \to (v'=2,L'=2)$ which benefits from almost perfect cancellation of the quadrupole shift (about $5 \times 10^{-17}$ relative to the transition frequency). Here, nonadiabatic calculations of the quadrupole moments were performed to get sufficient accuracy on the differential shift since the results of~\cite{Bakalov2013} are limited by the Born-Oppenheimer approximation. On that transition, it could be possible to reach a relative accuracy of about $6 \times 10^{-17}$. The blackbody radiation is not a limiting factor at this level: at 300K it represents about 10$^{-16}$ relative to the frequency, and the uncertainty could be reduced further by at least one order of magnitude by a careful determination of the environment temperature. The second-order Doppler effect is more likely to play a role, since it amounts to a few 10$^{-17}$ in the Al$^{+}$/Be$^+$ clock~\cite{Rosenband2008} and could become larger in the H$_2^+$/Be$^+$ case where the mass ratio is less favorable.

\begin{sidewaystable}
\begin{tiny}
\begin{tabular}{|c|c|c|c|c|c|c|c|c|c|c|c|c|c|c|c|c|}
 \hline
$(v,L)$ & $(v',L')$ &    $f_0$    & $f_0$ & $\Theta_{v,L,v',L'}^{(2)}$ & $|\Theta^{if}_{\epsilon,n}|$ & $\alpha_{v,L}^{(0)}$ & $\alpha_{v,L}^{(2)}$ & $\alpha_i$ & $\alpha_{v',L'}^{(0)}$ & $\alpha_{v',L'}^{(2)}$ & $\alpha_f$ & $\Delta\alpha$ & $\Delta f_{LS}$ & $\Delta f_Z$ & $\Delta f_Q$
     \\
&                   & [cm$^{-1}$] & [THz] &     [a.u.]            &   [a.u.]   &       [a.u.]         &        [a.u.]        & [a.u.]     &    [a.u.]              &     [a.u.]             &   [a.u.]   &      [a.u.]    &     [Hz]        &    [Hz]      &   [Hz]
     \\
 \hline
(0,1) & (1,1) & 2188.0613 & 65.60  & 3.762(-1) & 1.145(-2) & 3.1798 & -1.6950 & 3.0458 & 3.9129  & -2.4149  & 3.7220  & 0.6762 & 3.7(-3) & $\pm$7.0  & 1.1
\\
(0,2) & (1,2) & 2181.9499 & 65.41  & 4.118(-1) & 1.641(-2) & 3.1990 & -1.4465 & 3.0057 & 3.9382  & -2.0612  & 3.6628  & 0.6571 & 1.8(-3) & $\pm$14   & 1.5
\\
(0,3) & (1,1) & 1899.2023 & 56.94  & 5.295(-1) & 2.723(-2) & 3.2276 & -1.4176 & 3.6851 & 3.9121  & -2.4141  & 4.2938  & 0.6087 & 7.8(-4) & $\pm$1408 & -4.0
\\
(0,3) & (1,3) & 2172.8311 & 65.14  & 4.731(-1) & 1.924(-2) & 3.2280 & -1.4179 & 2.9992 & 3.9762  & -2.0207  & 3.6501  & 0.6509 & 1.3(-3) & $\pm$21   & 1.8
\\
(0,4) & (1,2) & 1780.7307 & 53.38  & 6.678(-1) & 3.029(-2) & 3.2663 & -1.4260 & 3.7750 & 3.9372  & -2.0603  & 4.4878  & 0.7128 & 8.4(-4) & $\pm$1413 & -1.6
\\
\hline
(0,1) & (2,1) & 4249.0149 & 127.38 & 2.887(-2) & 8.786(-4) & 3.1842 & -1.6992 & 3.0499 & 4.8583  & -3.3940  & 4.5900 & 1.5401 & 3.8(-1) & $\pm$15   & 2.2
\\
(0,2) & (2,2) & 4237.0011 & 127.02 & 3.169(-2) & 1.262(-3) & 3.2035 & -1.4501 & 3.0097 & 4.8927  & -2.8994  & 4.5053 & 1.4955 & 1.8(-1) & $\pm$30   & 3.1
\\
(0,3) & (2,1) & 3960.1559 & 118.72 & 2.776(-2) & 1.428(-3) & 3.2318 & -1.4208 & 3.6904 & 4.8555  & -3.3907  & 5.3916 & 1.7013 & 1.8(-1) & $\pm$1416 & -2.9
\\
(0,3) & (2,3) & 4219.0738 & 126.48 & 3.653(-2) & 1.485(-3) & 3.2325 & -1.4214 & 3.0031 & 4.9436  & -2.8435  & 4.4847 & 1.4816 & 1.3(-1) & $\pm$45   & 3.7
\\
(0,4) & (2,2) & 3835.7819 & 114.99 & 2.889(-2) & 1.310(-3) & 3.2704 & -1.4292 & 3.7803 & 4.8887  & -2.8956  & 5.6626 & 1.8823 & 2.6(-1) & $\pm$1428 & 0.006
\\
\hline
(0,1) & (3,1) & 6187.0684 & 185.48 & 4.044(-3) & 2.461(-4) & 3.1910 & -1.7057 & 3.0562 & 6.1156  & -4.7609  & 5.7392 & 2.6831 & 1.6(+1) & $\pm$23   & 3.4
\\
(0,2) & (3,2) & 6169.3302 & 184.95 & 4.448(-3) & 3.571(-4) & 3.2103 & -1.4556 & 3.0158 & 6.1627  & -4.0692  & 5.6189 & 2.6031 & 7.4     & $\pm$46   & 4.9
\\
(0,3) & (3,1) & 5898.2094 & 176.82 & 2.632(-3) & 0.3048    & 3.2384 & -1.4260 & 3.6986 & 6.1075  & -4.7514  & 6.8588 & 3.1601 & 1.7(+1) & $\pm$1424 & -1.7
\\
(0,3) & (3,3) & 6142.8586 & 184.16 & 5.146(-3) & 4.185(-4) & 3.2394 & -1.4268 & 3.0092 & 6.2324  & -3.9929  & 5.5880 & 2.5789 & 5.4     & $\pm$70   & 5.7
\\
(0,4) & (3,2) & 5768.1110 & 172.92 & 2.013(-3) & 0.3313    & 3.2770 & -1.4343 & 3.7887 & 6.1515  & -4.0579  & 7.2360 & 3.4473 & 4.3(+1) & $\pm$1445 & 1.8
\\
\hline
\end{tabular}
\caption{\label{lsquad} Same as Table~\ref{lsh2p}, for quadrupole ro-vibrational transitions between stretched states $i=(v,L,F=3/2,J=L+2,J_z=\pm(L+2)) \to f=(v',L',F'=1,S'=2,J'=L'+2,J'_z=\pm(L'+2))$ in H$_2^+$. $\Theta_{v,L,v',L'}^{(2)}$ is the reduced matrix element of the quadrupole operator between initial and final rovibrational states (see Sec.~\ref{sec-quad}). $|\Theta^{if}_{\epsilon,n}|$ is the amplitude of the hyperfine component that connects the stretched states $i,f$, taking into account the geometrical factors associated with the propagation direction and polarization of the probe laser (Eqs.~(\ref{quad-with-polar}) and (\ref{quad-exp})). The quadrupole shift $\Delta f_Q$ is calculated from the quadrupole moments published in~\cite{Bakalov2013} (with a corrigendum to be published), excepted for the $(v=0,L=4) \to (v'=2,L'=2)$ transition where a nonadiabatic calculation was performed to obtain a more precise estimate.}
\end{tiny}
\end{sidewaystable}

\section{Conclusion}

Our analysis of light shifts in high-resolution spectroscopy of hydrogen molecular ions has allowed assessing the performances of different types of transitions for a molecular ion clock that would set new limits on the variations of $\mu$. The lightshift was shown to be the main limiting factor in the case of two-photon transitions, both in H$_2^+$ and HD$^+$, it seems however possible to reach accuracy levels close to $5 \times 10^{-16}$ in a few cases. Quadrupole transitions are even more promising in view of the smaller required intensity; one particular transition that could allow going below the $10^{-16}$ level was pointed out. Hydrogen molecular ions thus have potential to improve the currently best laboratory limits on $\mu$-variation by several orders of magnitude. Finally, the lightshift can, in principle, be reduced by using a longer interrogation time, which would make the use of $v=0 \to v=3$ quadrupole transitions a realistic possibility. One advantage of such transitions is that they lye in a convenient wavelength range ($\lambda \sim 1.6-1.7$ $\mu$m) for laser technology and absolute frequency measurements. Of course, there are several experimental challenges on the way towards realization of molecular ion clocks, one of which being the preparation of the ions in the required hyperfine and Zeeman sublevel.

\vspace{5mm}

{\bf Acknowledgements.} I am grateful to V.I. Korobov for sharing his program for variational calculations of three-body systems, B. Dailly for his work on lightshift calculations, and L. Hilico and D. Bakalov for helpful discussions.

\section*{Appendix}

In this Appendix the methods used to compute the two-photon and quadrupole matrix elements including the effect of the hyperfine structure are briefly summarized.

\subsection*{1. Nonadiabatic wave functions}

To compute the wave functions of ro-vibrational states, the Schr\"odinger equation of the three-body Coulomb system is solved using a variational approach with exponential basis functions developed by V. Korobov~\cite{Korobov2000}. The wave function for a state with total orbital angular momentum $L$ and of a total spatial parity $\pi=(-1)^L$ is expanded as follows:
\begin{equation}\label{exp_main}
\begin{array}{@{}l}
\displaystyle \Psi_{LM}^\pi(\mathbf{r}_1,\mathbf{r}_2) =
       \sum_{l_1+l_2=L}
         \mathcal{Y}^{l_1l_2}_{LM}(\hat{\mathbf{r}}_1,\hat{\mathbf{r}}_2)
         G^{L\pi}_{l_1l_2}(r_1,r_2,r_{12}),
\\[4mm]\displaystyle
G_{l_1l_2}^{L\pi}(r_1,r_2,r_{12}) = \sum_{n=1}^N \Big\{C_n\,\mbox{Re} \bigl[e^{-\alpha_n r_{12}-\beta_n r_1-\gamma_n r_2}\bigr]
+D_n\,\mbox{Im} \bigl[e^{-\alpha_n r_{12}-\beta_n r_1-\gamma_n r_2}\bigr] \Big\}.
\end{array}
\end{equation}
where $r_1,r_2,r_{12}$ are the interparticle distances, and the complex exponents $\alpha$, $\beta$, $\gamma$ are generated in a pseudorandom way. Since very high accuracy is not required for transition probabilities, relatively small basis lengths of $N = 1000-2000$ were used, yielding a relative accuracy of at least a few parts in $10^{9}$ for the nonrelativistic energies, and a few parts in $10^{5}$ for the matrix elements.

\subsection*{2. Two-photon matrix elements}

The two-photon operator $^SQ$ defined in Eq.~(\ref{two-photon-op}), being a symmetrical tensor obtained by coupling of two vector operators, has a representation in terms of irreducible tensors that involves a scalar tensor $Q^{(0)a}$ and a tensor $Q^{(2)a}$ of rank 2. Its components $^SQ_{q_1,q_2}$ (where $q_i = -1,0,1$ correspond to the standard polarizations $\sigma^- , \pi , \sigma^+$) can be expressed in terms of the components of these irreducible tensors:
\begin{equation}
^SQ_{q_1 q_2}=\sum_{q=-2}^{2} a^{(2)}_q  Q^{(2)a}_q + a^{(0)}_0 Q^{(0)a}_0
\end{equation}
The values of the coefficients $a^{(k)}_q$ for all combinations of the standard polarizations can be found in Table~IV of~\cite{Karr2008a}. Here, we have adopted a slightly different definition for the irreducible tensors:
\begin{equation}
Q^{(0)} = - Q^{(0)a}/\sqrt{3}, \hspace{3mm} Q^{(2)} = \sqrt{2/3} \; Q^{(2)a}
\end{equation}
so that the linear-linear polarization component is simply written as:
\begin{equation}
^SQ_{00} = Q_{zz} = Q^{(0)}_0 + Q^{(2)}_0. \label{polar-lin-lin}
\end{equation}
Other components used in transition amplitude and lightshift calculations are:
\begin{eqnarray}
^SQ_{\pm 1 \pm 1} &=& \sqrt{3/2} \; Q^{(2)}_2 \label{polar-plus-plus} \\
-^SQ_{\pm 1 \mp 1} &=&  Q^{(0)}_0 - Q^{(2)}_0/2 \label{polar-plus-minus}
\end{eqnarray}

The first step is to calculate the reduced matrix elements $\left\langle vL\|Q^{(k)}\|v'L'\right\rangle$ with $k=0,2$. More precisely, the quantities given in the Tables are
\begin{equation}
Q_{v,L,v',L'}^{(k)} = \frac{\left\langle vL\|Q^{(k)}\|v'L'\right\rangle}{\sqrt{2 L + 1}} \label{two-photon-def}
\end{equation}
The calculation of two-photon reduced matrix elements has been described in~\cite{Karr2008a}; for easier reference we recall the main formulas here. The following three terms, which correspond to the possible values $L\!-\!1$, $L\!+\!1$, $L$ for the angular momentum of the intermediate state, are evaluated numerically:
\begin{eqnarray}
a_-&=&-\sum_{v''}\frac{\left\langle vL\|d\|v''L\!-\!1\right\rangle
                       \left\langle v''L\!-\!1\|d\|v'L'\right\rangle}
                          {\sqrt{(2L\!+\!1)(2L'\!+\!1)}(\hbar\omega-E_{v''L\!-\!1})}\\
a_+&=&-\sum_{v''}\frac{\left\langle vL\|d\|v''L\!+\!1\right\rangle
                       \left\langle v''L\!+\!1||d||v'L'\right\rangle}
                       {\sqrt{(2L\!+\!1)(2L'\!+\!1)}(\hbar\omega-E_{v''L\!+\!1})}\\
a_0&=&\sum_{v''}\frac{\left\langle vL\|d\|v''L\right\rangle
                      \left\langle v''L\|d\|v'L'\right\rangle}
                      {\sqrt{(2L\!+\!1)(2L'\!+\!1)}(\hbar\omega-E_{v''L})}
\end{eqnarray}
where $E_{v'',L''}$ is the energy of the intermediate state $|v''L''\rangle$ and $\omega$ is the photon energy. The reduced matrix elements of $Q^{(k)}$ are related to $a_-$, $a_+$, $a_0$ in the following way:
\begin{eqnarray}
\frac{\left\langle \!v\!L\|Q^{(0)}\|v'\!L\!\right\rangle}{\sqrt{2L+1}} &=& \frac{1}{3}\>\bigl(a_-+a_0+a_+\bigr)\>,\\
\frac{\left\langle \!v\!L\|Q^{(2)}\|v'\!L\!-\!2\!\right\rangle}{\sqrt{2L+1}} &=& -\sqrt{\frac{2(2L\!-\!3)}{3(2L\!-\!1)}}\>a_-\>,\\
\frac{\left\langle \!v\!L\|Q^{(2)}\|v'\!L\!\right\rangle}{\sqrt{2L+1}} &=& -\frac{1}{3}\sqrt{(\!2\!L\!+\!3\!)\!(\!2\!L\!-\!1\!)L\!(\!L\!+\!1\!)}\! \left[\frac{a_-}{L\!(\!2\!L\!-\!1\!)}\! -\!\frac{a_0}{L\!(\!L\!+\!1\!)}\!+\! \frac{a_+}{(\!2\!L\!+\!3\!)\!(\!L\!+\!1\!)}\right],\\
\frac{\left\langle \!v\!L\|Q^{(2)}\|v'\!L\!+\!2\!\right\rangle}{\sqrt{2L+1}}&=&-\sqrt{\frac{2(2L\!+\!5)}{3(2L\!+\!3)}}\;a_+\>.
\end{eqnarray}

Standard angular algebra procedures are then used to obtain the matrix elements between the initial and final hyperfine states. The expressions are made simpler by the fact that we only consider the ''stretched states'' which are ''pure'' states of angular momentum coupling. In the case of H$_2^+$ one obtains
\begin{eqnarray}
\left\langle v,L,F,J=L+F,J_z=\pm J| Q^{(k)}_q | v',L',F,J'=L'+F,J'_z=\pm J' \right\rangle = \nonumber \\
 -  \sqrt{2 J' + 1} \left\langle J'k J'_z q | J J_z \right\rangle \; \left\{\begin{array}{ccc}
          L&k &L'\\
          J'&F&J
       \end{array}\right\}
       \left\langle vL\|Q^{(k)}\|v'L'\right\rangle \label{hfs-h2p}
\end{eqnarray}
where $F = 1/2$ (resp. $3/2$) for even (resp. odd) values of $L$. The expression for HD$^+$ is similar:
\begin{eqnarray}
\left\langle \!v\!,\!L\!,\!F\!=\!1\!,\!S\!=\!2\!,\!J\!=\!L\!+\!2\!,\!J_z\!=\!\pm(L\!+\!2)| Q^{(k)}_q | v'\!,\!L'\!,\!F\!=\!1\!,\!S\!=\!2\!,\!J'\!=\!L'\!+\!2\!,\!J'_z\!=\!\pm(L'\!+\!2)\! \right\rangle = \nonumber \\
 -  \sqrt{2 J' + 1} \left\langle J'k J'_z q | J J_z \right\rangle \; \left\{\begin{array}{ccc}
          L&k &L'\\
          L'+2&2&L+2
       \end{array}\right\}
       \left\langle vL\|Q^{(k)}\|v'L'\right\rangle. \label{hfs-hdp}
\end{eqnarray}
The incident light is assumed to be linearly polarized for $\Delta L = 0$ transitions where $\Delta J_z = 0$ and circularly polarized for $\Delta L = \pm 2$ transitions where $\Delta J_z = \pm 2$.

\subsection*{3. Quadrupole matrix elements} \label{sec-quad}

In order to calculate its matrix elements, the quadrupole operator~(\ref{quad-def}) of a three-body system with masses $m_i$ and charges $Z_i$ ($i=1,2,3$) is written in terms of irreducible tensor operators associated with the coordinates $\mathbf{r_1}$, $\mathbf{r_2}$ and $\mathbf{r_{12}}$:
\begin{eqnarray}
\Theta_{ij} &=& c_1 \left(r_{1i} r_{1j} \!-\! \frac{\delta_{ij}}{3} r_1^2 \right) \!+\! c_2 \left(r_{2i} r_{2j} \!-\! \frac{\delta_{ij}}{3} r_2^2 \right) \!+\! c_3 \left(r_{12i} r_{12j} \!-\! \frac{\delta_{ij}}{3} r_{12}^2 \right), \\
c_1 &=& \frac{(m_1+m_3)^2 Z_2 + m_2^2 (Z_1 + Z_3)}{m_T^2} \\
c_2 &=& \frac{(m_1+m_2)^2 Z_3 + m_3^2 (Z_1 + Z_2)}{m_T^2} \\
c_3 &=& \frac{ m_2 m_3 Z_1 - - m_3 (m_1 + m_3) Z_2 -m_2 (m_1 + m_2) Z_3}{m_T^2}
\end{eqnarray}
with $m_T = m_1 + m_2 + m_3$. Using this expression one obtains the reduced matrix elements $\left\langle v,L \| \Theta^{(2)} \| v',L' \right\rangle$ (denoted $\Theta_{v,L,v',L'}^{(2)}$ in Table~\ref{lsquad}). Matrix elements between stretched hyperfine states $\left\langle i | \Theta^{(2)}_q | f \right\rangle$ can then be obtained from Eq.~(\ref{hfs-h2p}) (with the replacement of $Q^{(k)}$ by $\Theta^{(2)}$). The quantity $\Theta^{if}_{\epsilon,n}$ defined in Eq.~(\ref{quad-with-polar}) which appears in the Rabi frequency~(\ref{quad-rabi}) can be expressed in terms of matrix elements of the standard components of the quadrupole operator:
\begin{equation}
\Theta^{if}_{\epsilon,n} = \frac{1}{3} \sum_{ij} \sum_{q=-2}^{2} \left\langle i | \Theta^{(2)}_q | f \right\rangle \; c_{ij}^{(q)} \epsilon_i n_j \label{quad-exp}
\end{equation}
where the expressions of the second rank tensors $c_{ij}^{(q)}$ can be found in~\cite{James1998}. The choice of polarization and propagation direction of the probe laser is made so as to maximize the transition amplitude:
\begin{itemize}
{\item for $L \to L$, $\Delta J_z = 0$ transitions, we choose a propagation direction $\mathbf{n}$ making an angle of $\pi/4$ with the $z$ axis, and polarization in the plane defined by $\mathbf{n}$ and the $z$ axis. The geometrical factor $|g_0| = |\sum_{ij} c_{ij}^{(0)} \epsilon_i n_j|$ is then equal to $1/2$.}
{\item for $L \to L \pm 2$, $\Delta J_z = \pm 2$ transitions, we choose a propagation direction orthogonal to the $z$ axis, and polarization orthogonal to the plane defined by $\mathbf{n}$ and the $z$ axis. The geometrical factor $|g_{\pm 2}| = |\sum_{ij} c_{ij}^{(\pm 2)} \epsilon_i n_j|$ is equal to $1/\sqrt{6}$.}
\end{itemize}

\bibliography{mybibfile}

\end{document}